\documentclass[11pt]{article}
\newcommand{\beq}{\begin{equation}} 
\newcommand{\eeq}{\end{equation}}
\newcommand{\beqs}{\begin{eqnarray}} 
\newcommand{\eeqs}{\end{eqnarray}}
\newcommand{\gs}{g_\mathrm{s}}
\def\slashed#1{{\ooalign{\hfil\hfil/\hfil\cr $#1$}}}
\begin{document}
\begin{titlepage}
\vskip 2.5cm
\begin{center}
{\LARGE \bf Freezing of Moduli with Fluxes}\\
\smallskip
{\LARGE \bf in Three Dimensions}\\
\vspace{2.71cm}
{\Large
Riccardo Argurio,
Vanicson L. Campos, \\ 
\smallskip
Gabriele Ferretti and 
Rainer Heise}
\vskip 0.7cm
{\large \it Institute for Theoretical Physics - G\"oteborg University and \\
\smallskip
Chalmers University of Technology, 412 96 G\"oteborg, Sweden}
\vskip 0.3cm
\end{center}
\vspace{3.14cm}
\begin{abstract}
We study warped compactifications to three dimensions, realized as
an orientifold of type IIA string theory on $T^7$. By turning on 3- and
4-form fluxes on the torus in a supersymmetric way, we generate a
potential for the moduli fields. We present various flux configurations 
with ${\cal N}=1,2,3,4,5,6$ supersymmetries and count the number of
moduli in each case. In particular, we show that there are ${\cal N}=1$
configurations where all but one of the moduli are frozen. 
\end{abstract}

\end{titlepage}

\section{Introduction}
Generic supersymmetric compactifications of string and M-theory
lead to a low-energy effective theory in the non-compact dimensions
which contains a number of moduli fields. These are massless
scalars whose expectation value at infinity parameterizes 
the vacua of the low-energy theory.
Typically, it would be interesting to find compactifications which lead
to no moduli, for two main reasons. One is that in our four dimensional
world (or its minimal supersymmetric extension) we do not 
have any massless scalar field. The other reason is that a compactification
without moduli would give us a direct relation between string theory
and low-energy parameters, that is no tuning of the compactification
to fit low-energy data. This is of course desirable if the ultimate aim is
to derive low-energy physics entirely from first principles.

Of course, finding a string vacuum with no moduli is not an easy task. 
Recently, an interesting avenue for 
obtaining compactifications with a reduced number of moduli has been  
considered. It consists in turning on fluxes of the form field strengths
along the compactification manifold \cite{Polchinski}--\cite{Louis}
(see \cite{island} for a different approach  
based on asymmetric orbifolds yielding very few moduli). Generically, one
can give a number of conditions on the fluxes which imply that
the configuration preserves a certain amount of supersymmetry 
in the non-compact dimensions. Additionally,  these supersymmetric 
fluxes contribute to a tadpole cancellation condition, 
so that either manifolds with non-trivial
topology or orientifolds have to be considered.
Another interesting feature of these models is that they provide
a warp factor which implies that the scale in the non-compact dimensions
depends on the position in the compact manifold. This can potentially
lead to hierarchies in the low-energy theory (see e.g. 
\cite{Chan,Greene,Mayr,Giddings}).

If one considers a vacuum where supersymmetry is totally broken, then
most moduli will generically be massive at tree level, and the others
are likely to acquire masses by loop corrections. However, there is also
the danger that some of them become actually tachyonic or have a
potential unbounded from below.
We thus prefer to consider supersymmetric configurations where these
problems do not arise.

In \cite{Kachru}, a simple though illuminating example of moduli
stabilization through fluxes is considered. It consists of an orientifold
of type IIB theory on $T^6$, with 3-form fluxes turned on to compensate
(at least partially) for the O3-charge tadpole, while keeping ${\cal N}=1$ 
supersymmetry in 4 dimensions. It is shown that in the ``best'' case 
scenario, when all the O3-charge is compensated by fluxes (the addition of any
D3-brane brings in the moduli relative to its position), one is left
with 3 complex moduli, including most notably the volume of the torus.
Though there remain some moduli, it is still a drastic reduction, 
since one started from the 19 complex closed 
string moduli together with the 48 complex moduli corresponding to
the position of the 16 D3-branes present in the configuration without
fluxes.

In this note, we consider a 3 dimensional orientifold model. 
By going to a lower dimensional theory, we have
the possibility to consider a configuration with only 2 supercharges,
that is ${\cal N}=1$ in 3 dimensions. Lower supersymmetry implies less 
constraints on the low-energy Lagrangian, so that we expected to have
fewer moduli than in the case considered in \cite{Kachru}. 
In particular, we find models in which all but one of the moduli
are frozen.

Our model is built as follows. We consider an orientifold of type IIA theory 
on $T^7$. In order to cancel the tadpole generated by the O2-planes, 
we turn on the most general configuration of fluxes which preserves
at least the minimal amount of supersymmetry in 3 dimensions. 
If the fluxes alone cancel the O2-charge, the only potential moduli present
are the ones arising from closed strings. 
Without fluxes there are 64 of them. 
Some of them actually arise as massless abelian vectors, but in 3 dimensions
these are  Hodge dual to massless scalars.  
We will show that turning on the fluxes, many of the moduli (in the best
case all but one) are
frozen. More specifically, the metric moduli and the other scalars
are fixed by the condition for the fluxes to be supersymmetric. 
However, a combination of the dilaton and the volume of the torus is always a
free modulus.
The remaining 28 moduli would arise by dualizing the massless vectors
in the NSNS and RR sectors. However most of them (in the best case all of
them) acquire a mass through the Chern-Simons terms present in their
equations of motion, and thus cannot be dualized.
An interesting consequence of this effect is that in the generic case
the topological mass matrix for these fields is such that  the theory
breaks parity.

The outline of the paper is as follows. In section 2 we present
the orientifold and derive the fields which are present after 
the projection, including the allowed fluxes. In section 3 we consider
the action of type IIA supergravity, and how the requirement of
preserving supersymmetry in three dimensions leads to
constraints on the fluxes that can be turned on. In section 4 we
consider the tadpole cancellation and the Dirac quantization condition,
and show how the moduli are frozen, giving a number of explicit examples,
including an ${\cal N}=1$ example
where all the moduli but one acquire a mass.
In section 5 we discuss the 11 dimensional alternative interpretation
of our construction.
We conclude with some additional discussions.

\section{Type IIA on a $T^7/{\bf Z}_2$ orientifold}

Let us begin by reviewing the construction of type 
IIA string theory on a $T^7/{\bf Z}_2$ orientifold.

Contrary to type IIB string theory, where world-sheet parity $\Omega$ and
space-time left fermion number $(-1)^{F_L}$ are separately symmetries of
the theory, in type IIA, only $(-1)^{F_L}$ is a symmetry because
$\Omega$ would map spinors with opposite chirality into each other.
However, $\Omega$ can be combined with a parity transformation, that is the
reflection of an odd number of space coordinates, to yield a symmetry because
parity is implemented on the fermions by an odd number of gamma matrices.
We shall consider the generator of the ${\bf Z}_2$ orientifold group to be
$\Omega I_7$, where $I_7$ is the reflection of the seven coordinates 
of the torus. Our notation will be $M=(\mu, i)$ where $\mu =0,1,2$ are the 
uncompactified coordinates of three dimensional Minkowski space-time and 
$i=3, \cdots, 9$ the coordinates on the torus.

We will be using a mostly plus metric for which it is possible to choose
the ten-dimensional gamma matrices $\Gamma^M$ to be real
($\Gamma^{0T}=-\Gamma^0$) and define the action of
$\Omega I_7$ on the right- and left-moving spin-fields $S_\pm$:
\beq
    \Omega I_7\; : S_\pm \to PS_\mp\; , \label{oi}
\eeq
where the index $\pm$ denotes the ten-dimensional chirality:
$\Gamma^{11}S_\pm = \pm S_\pm$.
Of the two possible choices for the $P$-matrix
($P = \Gamma^{012}$ and
$P = \Gamma^{012}\Gamma^{11}\equiv \Gamma^{3456789}$) we choose the first 
because it squares to $+1$
thus ensuring that $(\Omega I_7)^2 = 1$ on the supercharges. Equivalently,
we could have chosen the second one and compensated for the minus sign by
the addition of $(-1)^{F_L}$. 

We can now study the ``parity'' of the various massless bosonic fields by
computing how they transform under $\Omega I_7$. For the NSNS fields 
$\Phi$, $G_{MN}$ and $B_{MN}$ the
analysis is the same as in the case of the IIB where:
\beqs
    \Omega I_7\; : \Phi &\to& \Phi \nonumber\\
                   G_{\mu\nu} &\to& G_{\mu\nu}  \nonumber \\
                   G_{\mu i} &\to& - G_{\mu i}  \nonumber \\
                   G_{i j} &\to& G_{i j}  \nonumber \\
                   B_{\mu\nu} &\to& -B_{\mu\nu}  \nonumber \\
                   B_{\mu i} &\to& B_{\mu i}  \nonumber \\
                   B_{i j} &\to& -B_{i j} \; . 
\eeqs
For the RR fields $A_M$ and $A_{MNP}$ we use the action of (\ref{oi}) on
the spin fields $S_\pm$ entering the definition of the vertex operators and
the following relation, (valid for any pair of ten-dimensional 
Majorana spinors $\epsilon$ and $\lambda$):
\beq
     \bar\epsilon \;\Gamma^{M_1\cdots M_n}\; \lambda = 
     (-1)^n \bar\lambda \; \Gamma^{M_n\cdots M_1}\; \epsilon
\eeq
to get
\beqs
    \Omega I_7\; : A_\mu &\to& - A_\mu \nonumber\\
                   A_i &\to& A_i\nonumber\\
                   A_{\mu\nu\rho} &\to& A_{\mu\nu\rho}\nonumber\\
                   A_{\mu\nu i} &\to& - A_{\mu\nu i}\nonumber\\
                   A_{\mu i j} &\to& A_{\mu i j}\nonumber\\
                   A_{i j k} &\to& - A_{i j k}\; . \label{rr}
\eeqs
As an example, consider the vertex operator for 
$F_{\mu i} = \partial_\mu A_i - \partial_i A_\mu$: \hfill\break 
$V \propto \bar S_- \; \Gamma^{\mu i}\; S_+$. Under the action of 
$\Omega I_7$:
\beq
    V \to S_+^T P^T \Gamma^0 \; \Gamma^{\mu i}\; P S_- =
          - \bar S_+ \; \Gamma^{\mu i}\;  S_- =
          - \bar S_- \; \Gamma^{i \mu}\;  S_+ = V.
\eeq
Thus, $\Omega I_7: F_{\mu i} \to F_{\mu i}$. The various 
transformation properties for the gauge potentials 
in (\ref{rr}) follow from these types of calculation taking into account the
extra minus sign coming from $\partial_i$ when going from $F$ to $A$.

We see that $F_{ijkl} = 4\partial_{[i}A_{jkl]}$ and 
$H_{ijk} = 3\partial_{[i}B_{jk]}$ have the right parity properties to be
constant fluxes but not $F_{ij} = 2\partial_{[i}A_{j]}$.

Now recall that, in three dimensions, the metric and a two-form or 
three-form gauge potential carry no degree of freedom, whereas a massless
vector field can be dualized into a scalar.
When counting the total number of scalars that are not projected out, 
we must count also those that come from dualizing a vector. 
The complete list is thus
\beqs
                   \Phi &:& \hbox{1 scalar} \nonumber\\
                   G_{i j} &:& \hbox{28 scalars} \nonumber \\
                   B_{\mu i} &:& \hbox{7 vectors} \nonumber \\
                   A_i  &:& \hbox{7 scalars} \nonumber\\
                   A_{\mu i j}  &:& \hbox{21 vectors}
\eeqs
for a total of 64 bosonic degrees of freedom.

\section{Type IIA dynamics}

The bosonic part of the action of type IIA supergravity, written
in the Einstein frame ($G_{MN}^\mathrm{Einst}=e^{-\phi /2}
G_{MN}^\mathrm{string}$), is the following \cite{IIA}:
\beqs
S&= {1\over (2\pi)^7 {\alpha'}^4}\int d^{10}x & \left\{  \sqrt{-G}
\left[ R -{1\over 2}\partial_M \phi \partial^M \phi
-{1\over 12} e^{-\phi} H_{MNP}H^{MNP} 
\right. \right. \nonumber \\
& & \left. -{1\over 4} e^{3\phi \over 2} F_{MN}F^{MN}
-{1\over 48} e^{\phi \over 2} \tilde{F}_{MNPQ}\tilde{F}^{MNPQ}\right]
\nonumber \\
& & \left. -{1\over (48)^2} 
\varepsilon^{MNPQRSTUVW}B_{MN}F_{PQRS}F_{TUVW}\right\}\ .
\label{action}
\eeqs
In the above equation we have defined:
\beq
\tilde{F}_{MNPQ}=F_{MNPQ} + 4 A_{[M}H_{NPQ]}. 
\label{ftilde}
\eeq
All the field strengths are defined as usual as 
$H_{MNP}=3\partial_{[M} B_{NP]}$ 
and similarly for the others, and we take the tensor
$\varepsilon^{MNPQRSTUVW}$ to be independent of the metric $G_{MN}$,
with $\varepsilon^{0123456789}=-1$.

The supersymmetry transformations of type IIA  
supergravity in a purely bosonic background are given 
by the dilatino and gravitino variations as follows \cite{IIA}:

\beqs
\delta\lambda &=& 
-D_M \phi \Gamma^M \Gamma^{11} \epsilon + \frac{3}{8} 
e^{\frac{3\phi}{4}}\Gamma^{MN}F_{MN}\epsilon
 \nonumber \\ && 
+ \frac{1}{12}e^{-\frac{\phi}{2}}
\Gamma^{MNL}H_{MNL} \epsilon 
+ \frac{1}{96} e^{\frac{\phi}{4}} \Gamma^{MNLP}\tilde F_{MNLP} 
\Gamma^{11}\epsilon
\, ,
\label{dilatino}
\eeqs
\beqs
\delta \Psi_M &=& D_M \epsilon - \frac{1}{64}
e^{\frac{3\phi}{4}}({{\Gamma}_{M}}^{NL} - 14 \delta^N_M \Gamma^L) 
F_{NL}\Gamma^{11}\epsilon
 \nonumber \\ && 
+ \frac{1}{96} e^{-\frac{\phi}{2}} 
({{\Gamma}_{M}}^{NLP} - 9{\delta}^N_M \Gamma^{LP})H_{NLP}\Gamma^{11} \epsilon
 \nonumber \\ &&
+ \frac{1}{256} e^{\frac{\phi}{4}} ({{\Gamma}_{M}}^{NLPQ} - 
\frac{20}{3} 
{\delta}^N_M \Gamma^{LPQ})\tilde F_{NLPQ}\epsilon \, ,
\label{gravitino}
\eeqs
where 
$\{ \Gamma^M,\Gamma^N \} = 2G^{MN}$, 
$\Gamma^{11}$ is independent of the metric and
the supersymmetry parameter $\epsilon$ is a real Majorana 32-component spinor.

To solve the above supersymmetry equations we take a D2-brane-like ansatz in
terms of the warp factor $\Delta$ as follows:
\beqs
ds^2 &=& \gs^{-1/2} \left(\Delta^{-\frac{5}{8}}\eta_{\mu\nu}dx^\mu dx^\nu 
+ \Delta^{\frac{3}{8}}g_{mn} dx^m dx^n\right) \nonumber \\
F_{(4)} &=& \gs^{-1} dx^0 \wedge dx^1 \wedge dx^2 \wedge d\Delta\, 
\Delta^{-2}, \qquad 
e^{\phi} = \gs \Delta^{\frac{1}{4}}.
\label{ansatz}
\eeqs
We also allow for non-zero fluxes $H_{ijk}$ and $F_{ijkl}$ and take the 
supersymmetry parameter to satisfy 
$\Gamma^{012}\epsilon = \epsilon$ (where here the indices are flat).

Then the vanishing of the dilatino equation (\ref{dilatino})
implies the relation:
\beq
\frac{1}{12}e^{-\frac{\phi}{2}} 
\Gamma^{mnl}\, H_{mnl} \epsilon  + \frac{1}{96}e^{\frac{\phi}{4}} 
\Gamma^{mnlp}\tilde{F}_{mnlp}\Gamma^{11}\epsilon = 0. \label{dila}
\eeq

The vanishing of the gravitino variation (\ref{gravitino}) 
along the $\mu$-direction implies instead (assuming, of course, 
$\partial_\mu\epsilon = 0$):
\beq 
\frac{1}{96}e^{-\frac{\phi}{2}} 
{{\Gamma}_{\mu}}^{mnp} H_{mnp} \Gamma^{11}\epsilon
+\frac{1}{256} e^{\frac{\phi}{4}}
{{\Gamma}_{\mu}}^{mnlp}\tilde{F}_{mnlp} \epsilon = 0.
\label{gravitinoMU}
\eeq

Equations (\ref{dila}) and  (\ref{gravitinoMU}) together with 
$\Gamma^{012}\epsilon = \epsilon$ require:
\beqs
\slashed{H}\epsilon &=& 0, \label{hsl}\\
\slashed{\tilde{F}}\epsilon &=& 0, \label{fsl}
\eeqs
where the slash represents the contraction of the fluxes with the 
gamma matrices.

Finally, we are left with the $\delta\Psi_m=0$ condition, which,
after a rescaling of the spinor $\epsilon \to \; \Delta^{-5/32}\epsilon$ gives:
\beq
\frac{1}{96}e^{-\frac{\phi}{2}}
({{\Gamma}_{m}}^{nlp}H_{nlp} - 9\, \Gamma^{nl}H_{mnl})
\Gamma^{11}\epsilon
+ \frac{1}{256}e^{\frac{\phi}{4}} ({{\Gamma}_{m}}^{nlpq}
\tilde{F}_{nlpq}
- \frac{20}{3}\Gamma^{nlp}\tilde{F}_{mnlp})\epsilon = 0
\eeq

This vanishes if we impose:
\beq
\tilde{F}_{mnlp} = 
\frac{1}{6}e^{-\frac{3 \phi}{4}} \sqrt{G_{(7)}}
\varepsilon_{mnlpqrs} H^{qrs}, \label{duality}
\eeq
where $G_{(7)}=\det (G_{mn})$.
After substituting the ansatz (\ref{ansatz}) 
(e.g. $G_{mn} = \gs^{-1/2} \Delta^{3/8} g_{mn}$), we see that
all the dependence on the warp factor $\Delta$ disappears
and we obtain:
\beq
F_{mnlp} + 4 A_{[m}H_{nlp]}= \frac{1}{6} \gs^{-1} \sqrt{g_{(7)}}\ 
\varepsilon_{mnlpqrs} H_{ijk}g^{qi}g^{rj}g^{sk}, \label{dualHF}
\eeq 
where the metric used is the same $g_{mn}$ as in 
(\ref{ansatz}), $g_{(7)}=\det(g_{mn})$, and we have used (\ref{ftilde}).

Thus, the extra conditions for supersymmetry due to the fluxes are
(\ref{hsl}) and (\ref{dualHF}), with equation (\ref{fsl})
following from the previous two. The fact that  
(\ref{hsl}) and (\ref{dualHF}) are necessary and sufficient conditions
for preservation of the minimal amount of supersymmetry has been 
shown in \cite{becker} in the context of 11 dimensional
supergravity on a 8-manifold, and the same holds true in our case after
trivial dimensional reduction.

It can be checked that the ansatz (\ref{ansatz}) is indeed a solution
to the equations of motion derived from (\ref{action}) provided the
relation (\ref{duality}) is realized. In particular, the equations
of motion of the form fields along the directions of the torus
would not vanish for an arbitrary configuration of fluxes.
The duality relation (\ref{duality}) was derived in this way in a 
related context in \cite{hawking} (see also \cite{cvetic}).

The remaining non-trivial equations all reduce to an equation
for the warp factor $\Delta$ which reads:
\beq
{1\over  \sqrt{g_{(7)}}}\partial_m (\sqrt{g_{(7)}}\ g^{mn}\partial_n \Delta) =
-{1\over 6} H_{mnp}H^{mnp} -(2\pi\sqrt{\alpha'})^5\rho_{D2-O2}.
\eeq
On the right-hand side of this equation we have included the contribution
from localized sources on the torus, that is the number density of
orientifold O2-planes and of
D2-branes which are possibly present. The integral on the torus
of the left-hand side must vanish, and thus the sum of the contributions
of the fluxes and the charges must vanish too. This is just the tadpole 
cancellation condition, since the above equation is also derived
from the equations of motion of the 3-form potential along the 012 directions.

Note that in order for the ansatz (\ref{ansatz}) to be a solution of the
equations of motion in the presence of non-vanishing fluxes, it is sufficient
to impose (\ref{duality}). Namely, a solution to the equations
of motion can violate (\ref{hsl}) and thus be non-supersymmetric.

\section{Generating mass terms for the moduli}

We have seen that the conditions required by supersymmetry on the fluxes
$H_{ijl}$ and $F_{ijkl}$ are (\ref{hsl}) and (\ref{dualHF}).

There are two more conditions that the fluxes must satisfy in order to 
have a consistent string vacuum. The first is the quantization condition 
which is expressed by:
\beq
     \frac{1}{2\pi\alpha^\prime}\int_C H = 2\pi m, \quad
     \frac{1}{(2\pi)^2\alpha^{\prime 3/2}}\int_{\tilde C} F = 2\pi n,
         \label{fluxquant}
\eeq
where $C$ and $\tilde C$ are respectively three-cycles and four-cycles
of $T^7$ and we shall always assume $m$ and $n$ to be \emph{even} to avoid
problems with the ``half-cycles'' discussed in \cite{Kachru}
and \cite{Frey}.

The second condition is that the total D2-charge induced by the fluxes
does not exceed (minus) the orientifold charge. If this condition is 
satisfied, 
D2-branes can then be used to make up for the difference in order to
cancel the RR tadpole, whereas, if the charge induced by the fluxes exceeded
the bound, anti D2-branes would have to be used, thus breaking supersymmetry.
In formulas, the condition is:
\beq
     \frac{1}{2} N_{\hbox{flux}} = \frac{1}{2} 
     \frac{1}{(2\pi)^5\alpha^{\prime 5/2}} \int_{T^7} H \wedge F \leq 16.
     \label{tadpole}
\eeq

Equations (\ref{fluxquant}, \ref{tadpole}) require that one writes:
\beqs
    H &=& \frac{1}{6} {H_{ijk}\over 2\pi\sqrt{\alpha'}} 
dx^i\wedge dx^j \wedge dx^k,
    \nonumber \\
    F &=& \frac{1}{24} {F_{ijkl} \over 2\pi\sqrt{\alpha'}}
       dx^i\wedge dx^j \wedge dx^k \wedge dx^l,
\eeqs
where $x^i \sim x^i+2\pi \sqrt{\alpha'}$, and 
we have rescaled the fields in such a way that now
$H_{ijk}$ and $F_{ijkl}$ are (even) integers.
Equation (\ref{tadpole}) then requires that those integers obey:
\beq
     \frac{1}{144}\varepsilon^{ijklmnp} H_{ijk} F_{lmnp} \leq 32,
     \label{simpletad}
\eeq
where the epsilon tensor is purely numerical.

\subsection{Scalars}

Let us now derive a potential for the candidate moduli. 
We will start with the massless scalars, 
that is the dilaton $\gs$, the 28 metric elements $g_{mn}$ and the
7 $A_i$ scalars.

Equation (\ref{dualHF}) constrains these moduli, but 
it is clear that the model always allows for a flat direction characterized
by the rescaling $g_{mn} \to \lambda g_{mn}$ and 
$\gs \to \lambda^{1/2} \gs$. This modulus will never be 
stabilized and we shall return to its physical interpretation later on.
We can thus write $g_{mn}=\gs^2 \hat g_{mn}$ so that $\gs$ is rescaled
away from the relation. It is now clear that we can aim at freezing
all of the $\hat g_{mn}$ and $A_i$ moduli.

The potential for these moduli comes from the kinetic terms for the
3- and 4-form field strengths in (\ref{action}). Up to an overall
numerical factor, it is given by:
\beq
V(\gs, \hat g_{ij}, A_i)={1\over \gs} \sqrt{\hat g} \left( {1\over 12}
H_{ijk} H_{mnp} \hat g^{im}\hat g^{jn}\hat g^{kp} +{1\over 48}
\tilde{F}_{ijkl} \tilde{F}_{mnpq}
\hat g^{im}\hat g^{jn}\hat g^{kp}\hat g^{lq}\right).
\label{potential}
\eeq
Recall that $\tilde{F}_{ijkl}$ depends on $A_i$ as in (\ref{ftilde}).

We now want to develop this potential around a solution and show that
it generically leads to masses for the $\hat g_{mn}$ and $A_i$ moduli.
However, finding the most general solution is a complicated task.
We shall bypass this problem by making an ansatz that captures the 
essential features of the model while allowing for a complete solution.
We shall consider a
point in moduli space where we \emph{know} that a solution with
integer fluxes exists, namely the square torus $\hat g_{ij} = \delta_{ij}$,
$A_i=0$ so that we can solve for $F_{ijkl}$ as a 
function of $H_{ijk}$:
\beq
    F_{ijkl} = \frac{1}{6}\varepsilon_{ijklmnp}H_{mnp}, \label{sqtorus}
\eeq
where, again, $\epsilon_{ijklmnp}$ is purely numerical and the repeated 
indices are summed with the square metric $\delta_{ij}$ so they will all
be written downstairs.

We can now expand the metric
to linear order $\hat g_{ij} = \delta_{ij} + h_{ij}$ and rewrite the
potential (\ref{potential}) taking into account (\ref{sqtorus}):
\beqs
\gs V(h_{ij})& = & {1\over 6} H_{ijk}H_{ijk} \nonumber \\
   & + & {1\over 48}  h_{ii} h_{jj} H_{klm} H_{klm} 
-{1\over 4} h_{ii} h_{jk} H_{jlm}H_{klm}  \nonumber \\
 & +& {1\over 4} h_{ij} h_{jk} H_{ilm}H_{klm} 
+ {1\over 2} h_{ij} h_{kl} H_{ikm}H_{jlm}\nonumber \\
&+& {1\over 2}h_{ij}A_k F_{iklm}H_{jlm}\nonumber \\
&+& {1\over 12} A_i A_i H_{jkl}H_{jkl} -{1\over 4}A_i A_j  H_{ikl}H_{jkl}
+\dots .  \label{quadpot}
\eeqs
The zeroth order term is actually exactly canceled by the (negative)
contribution of the orientifold tension due to the tadpole condition.
We note that $\hat g_{ij} = \delta_{ij}$, $A_i=0$ is indeed an extremum
of the potential. The next step would be to diagonalize the mass matrix
derived from the second order term above and count the number of zero
eigenvalues.

An alternative way to see how many moduli 
are frozen is to investigate the space of
solutions of the linearized form of eq.~(\ref{dualHF}), after taking into
account (\ref{sqtorus}):
\beq
    h_{ii}H_{mnp} -2\left( h_{ip} H_{mni} +  h_{im} H_{npi} + 
    h_{in} H_{pmi} \right) +2A_iF_{imnp}= 0 \label{line} .
\eeq
It can be checked that the quadratic potential (\ref{quadpot}) is 
proportional to the square of (\ref{line}). This is in keeping with
the interpretation of the duality equation (\ref{dualHF}) as coming
from the variation of a superpotential. 
It also implies that all the non-zero eigenvalues of the mass matrix
are positive, even for non-supersymmetric configurations.
Eq.~(\ref{line}) is easier to
handle and the problem of counting the number of massless moduli is
reduced to that of computing the rank of a 35$\times$35 matrix with
integer coefficients, which can be easily performed by a computer 
program. A condition for the stabilization of all these moduli is 
that the system admits only the trivial solution $h_{ij} = 0$ and $A_i=0$.

Inserting (\ref{sqtorus}) into (\ref{simpletad}) we also obtain: 
\beq
\frac{1}{6} H_{ijk}H_{ijk} \leq 32. \label{hsq}
\eeq
If we want to saturate the inequality (\ref{hsq}),
thus allowing for no D2-branes, we have the following
three possibilities:
\begin{itemize}
\item Eight non-vanishing values for $H_{ijk}$ all equal to $2$;
\item One entry equal to $4$ and four equal to $2$;
\item Two non-vanishing values for $H_{ijk}$ both equal to $4$.
\end{itemize}

Before discussing particular solutions, let us first consider
also the massless vectors.

\subsection{Vectors}
The remaining moduli fields would arise as Hodge duals
of the 3 dimensional vectors $B_{\mu i}$ and $A_{\mu ij}$. Without
fluxes, each one of these fields appears with no interactions
whatsoever and can thus be dualized to a massless scalar.
In the presence of fluxes, the Chern-Simons (CS) term of the 10 dimensional
action will be non-vanishing and this will provide topological 
mass terms to the 28 vector fields \cite{templeton}. 

The correct 3 dimensional CS term is better derived 
through the reduction of the equations of motion, since in the action
(\ref{action}) the CS term is actually defined up to a total derivative
term which is relevant here. After taking that into account, one
finds the following 3 dimensional action:
\beqs
S &=\int d^3x & \left[-{1\over 4}H_{\mu\nu i}{H^{\mu\nu}}_i
-{1\over 8}F_{\mu \nu ij}{F^{ \mu \nu}}_{ij} \right. \nonumber \\
& & \left. -{1\over 4}\varepsilon^{\mu
\nu\rho}\left(B_{\mu i}F_{\nu \rho jk}H_{ijk}
+{1\over 4} A_{\mu ij}F_{\nu \rho kl}F_{ijkl}\right)\right],
\eeqs
where we have used the relation (\ref{sqtorus}). 
Note that when the sum is performed
over the torus indices all the kinetic terms acquire the 
canonical normalization and the CS terms have the same numerical prefactor.

Here we notice that the CS term couples together the 28 vector fields
through a matrix which is symmetric in the above fields. It is then
possible to diagonalize it by an orthogonal matrix, and every non-vanishing 
eigenvalue will lead to a topological mass for the relative vector field
\cite{templeton}. Such a topologically massive vector is impossible
to dualize to a massless scalar, and hence these fields do not lead to moduli.
It is interesting to note that quite generically the eigenvalues
of the topological mass matrix break parity. For instance, for a positive
eigenvalue there will not necessarily be a negative eigenvalue of the
same magnitude.

\subsection{Some examples with ${\cal N}$=1,2,3,4,5,6}
Let us now present various configurations of fluxes leading to all possible
amounts of supersymmetry allowed by the problem. 
For each case we will comment on the number of
moduli. We consider only cases where the tadpole condition is saturated,
that is one of the three cases listed in subsection 4.1. Although
a complete classification of the different results obtained by specific
choices of the $H_{ijk}$ is beyond the scope of this work, it is 
interesting to see some of the possible outcomes of the procedure
of turning on supersymmetric fluxes.

In the following we will decompose the ten dimensional flat gamma matrices as 
follows:
\beqs
    \Gamma^\mu &=& \gamma^\mu \otimes {\bf 1} \otimes \sigma_1 
                                               \nonumber \\
    \Gamma^i &=& {\bf 1} \otimes \hat\gamma^i \otimes \sigma_2
\eeqs
and consequently, the $32$ component spinor $\epsilon$ as a $2\times 8\times 2$
spinor $\epsilon = \chi\otimes\eta\otimes\alpha$.
The torus gamma matrices are $8\times 8$ purely imaginary and hermitian.

Supersymmetric configurations (\ref{hsl})
can be easily obtained by deriving the 
equations that preserve a particular spinor $\eta$, that is
by writing $H_{ijk}\hat\gamma^{ijk}\eta=0$ for a specifically chosen 
$\eta$ and choosing $H_{ijk}$ so that the equations are satisfied. 

In this section we relabel the coordinate on the torus 
from $3\cdots 9$ to $1\cdots 7$ since there is no possibility of confusion.

\subsubsection*{${\cal N}=1$}
Let us start with a configuration of fluxes which is the most effective
in lifting moduli, i.e. where only one modulus is left. 
An example of ${\cal N}=1$ solution 
freezing the 35 scalars $\hat{g}_{ij}$ and $A_i$ is:
\beq
H_{123} = H_{127} = H_{136}= H_{235}= H_{236} = H_{257}= H_{347}= H_{357}= 2.
\label{hsolution}
\eeq
Similarly, one can check that there are no zero eigenvalues to the
CS topological mass matrix and consequently
there are no moduli arising from the 28 vector fields.

Other cases with ${\cal N}=1$ can be found, but many of them will have
more than one modulus. One such example is given by:
\beq
H_{127} =4, \quad H_{126}= H_{134}= H_{156} = H_{247}= 2,
\label{hsolution2}
\eeq
which gives rise to 4 moduli from the scalars, while it gives masses
to all the vectors.

\subsubsection*{${\cal N}=2$}
An ${\cal N}=2$ configuration can at most freeze all but 2 moduli,
since the dilaton/volume has to sit in a supermultiplet, which
in turn has to have two bosonic degrees of freedom (for a table
of the dimensions of massless supermultiplets in 3 dimensions with
various supersymmetries, see \cite{deWit}). One such configuration
is given by:
\beq
H_{126} = H_{127} = H_{134}= H_{135}= H_{234} = H_{236}= H_{247}= H_{357}= 2.
\label{hsolution3}
\eeq
The other modulus in the massless supermultiplet with the dilaton/volume
arises from the vector fields, since the CS mass matrix has one zero 
eigenvalue. Note that in this case and all the following ones, the three
dimensional graviton
multiplet never gives rise to moduli since all its components
are purely topological.

Another class of flux configurations that must give
${\cal N}=2$ supersymmetries in 3 dimensions is the one obtained
by trivial dimensional reduction and T-duality
from the minimally supersymmetric configurations of \cite{Kachru}. 
Under transverse T-duality the $H_{ijk}$ is unaffected and thus we
can simply consider solutions where the  $H_{ijk}$ fluxes never
wrap, say, the 7th direction of the torus. The minimal
number of moduli we expect
is thus 8, since in 4 dimension \cite{Kachru} there were 3 complex moduli
together with the graviton multiplet, which upon dimensional reduction
gives 2 more massless bosonic degrees of freedom. Indeed, the example:
\beq
H_{123} = H_{126} = H_{135}= H_{146}= H_{156} = H_{246}= H_{345}= H_{456}= 2,
\label{hsolution4}
\eeq
has 4 massless scalars and 4 massless vectors.

\subsubsection*{${\cal N}=3$}
More exotic is the case with ${\cal N}=3$ supersymmetries. Here the massless
multiplets have dimension 4, so we expect the moduli to come in that 
multiplicity. An example with only 1 massless supermultiplet is:
\beq
H_{127} = H_{134} = H_{156}= -H_{235}= H_{236} = H_{245}= H_{357}= -H_{467}= 2,
\label{hsolution5}
\eeq
with the dilaton/volume modulus and 3 massless vectors.

There are many more ${\cal N}=3$ cases with more than 1 massless 
supermultiplet, for instance:
\beq
H_{123} = H_{126} = H_{135}= H_{156}= H_{234} = H_{246}= H_{257}= H_{456}= 2,
\label{hsolution6}
\eeq
with 3 supermultiplets giving a total of
7 massless scalars and 5 massless vectors.

\subsubsection*{${\cal N}=4$}
${\cal N}=4$ supersymmetries are the generic case when only two fluxes
are turned on in a supersymmetric way and saturate the tadpole. In fact, 
all such cases are equivalent, since the condition
$\slashed{H}\eta = 0$ has non-zero solutions if and only if the two fluxes
have only one direction in common. Furthermore, this will induce a projection
which preserves exactly half of the components of the spinor, thus giving
${\cal N}=4$. We can thus take:
\beq
H_{123} = H_{345} = 4,
\label{hsolution7}
\eeq
and find that this class of configurations gives 14 massless moduli and
10 massless vectors, for a total of 6 ${\cal N}=4$ multiplets.

For ${\cal N}=4$ configurations with 8 fluxes turned on, 
we can find a different number of moduli. For instance, for:
\beq
H_{126} = H_{135} = H_{136}= H_{145}= H_{247} = H_{256}= H_{347}= H_{567}= 2,
\label{hsolution8}
\eeq
we find 8 massless scalars and 8 massless vectors.
Actually, all the other ${\cal N}=4$ configurations we have tried
have the same set of moduli as one of the two configurations above.

\subsubsection*{${\cal N}=5$}
We have also found ${\cal N}=5$ configurations, like:
\beq
H_{134} =-H_{156} = H_{236}=-H_{245}=-H_{347} = H_{357}= H_{467}= H_{567}= 2,
\label{hsolution9}
\eeq
with 6 massless scalars and 10 massless vectors. All the other 
${\cal N}=5$ configurations that we have found have the same set of moduli.
Note that here the massless supermultiplets are 8-dimensional \cite{deWit}.

\subsubsection*{${\cal N}=6$}
${\cal N}=6$ configurations can be derived from the 4 dimensional ${\cal N}=3$
cases of \cite{Frey}. For instance, for:
\beq
H_{123} = H_{126} = H_{135}= H_{156}= H_{234} = H_{246}= H_{345}= H_{456}= 2,
\label{hsolution10}
\eeq
we find 16 massless scalars and 16 massless vectors.

It is not possible to have ${\cal N}=7$ configurations, since that would
imply that $\slashed{H}$ has only one non-zero eigenvalue. 
However, this is impossible because $\slashed{H}$ is a traceless matrix.
Thus the minimum amount of non-zero eigenvalues is 2.
The only remaining case, ${\cal N}=8$, is the trivial one when the
fluxes are not turned on and all the 64 closed string moduli are massless
(moreover the tadpole has to be canceled by the addition of 16
D3-branes). 

The upshot of this review of particular configurations is that it is 
fairly easy to come up with a configuration of fluxes for any given
amount of supersymmetry which is not ruled out by general arguments.
Although for ${\cal N}=1,2,3$ it is possible to find configuration
with the minimal number of moduli, they do not seem to be generic.
Moreover for ${\cal N}=4,5,6$ we have not been able to find such 
configurations, hinting that for higher supersymmetry the low-energy
theories deriving from compactifications with fluxes are not generic.

\section{An 11-dimensional perspective}
As we have seen in the previous section, the modulus which is always
present is a simultaneous rescaling of the string coupling and of
the volume of the torus, in such a way that we can write
$g_{ij}=\gs^2\delta_{ij}$. This means that the radii of the directions
of the torus are given by $R_i=\gs \sqrt{\alpha'}$. 
On the other hand, when uplifting a type IIA configuration to M-theory, 
one gets the same relation for the radius of the 11th direction, 
namely $R_{11}=\gs \sqrt{\alpha'}$.
Thus the correct interpretation is that the modulus is the volume
of the 8-dimensional torus in the M-theory picture of the configuration.

In fact, it is possible to do all the analysis in this paper from
an M-theory perspective. One starts by considering M-theory on
a $T^8/{\bf Z}_2$ orbifold which is the uplift of the configuration
of type IIA theory with the O2-planes which we have considered here
(see e.g.~\cite{sen,sethi,hanany}). Then, upon implementing the M2-brane
ansatz together with fluxes on $T^8$ and requiring supersymmetry, 
one obtains a condition similar to (\ref{fsl}) supplemented by a
a relation of self-duality on the torus for the 4-form field strength
fluxes. The fluxes now have to saturate a tadpole condition coming
from the Euler number of the orbifold which obviously matches the total
O2-plane charge in 10-dimensions. The analysis of the potential
for the metric moduli and the CS-terms for the vectors goes along
very similar lines as in the previous section.
The two approaches are clearly exactly equivalent.

We should also consider the order of magnitude of the 
masses that are generated through the presence of fluxes. It can be shown
that all the masses generated by the fluxes, both the ones of the scalars
and the topological masses of the vectors, satisfy:
\beq
m \sim {\alpha' \over R^3}\ ,
\label{magn}
\eeq
where $R$ is the radius of any direction of the torus, as given by
the remaining modulus. Note that this mass can be also expressed
as $m \sim (\gs^3 \sqrt{\alpha'})^{-1}$. We note now that as long
as $R> \sqrt{\alpha'}$ (and thus $\gs >1$), the masses coming from
the fluxes are smaller than any other scale of any other massive object
in the theory, be they KK excitations on the torus, string massive
states or wrapped D-branes. It is thus consistent, in this regime, 
to consider a truncated low-energy theory comprising the states 
which are massive due to the fluxes \cite{Kachru}.
Note however that we are not allowed to take $\gs<1$, since this would
imply also $R< \sqrt{\alpha'}$ and the breakdown of the low-energy 
description in terms of supergravity (we could however still consider
the effective theory of the massless modes).

The fact that $\gs >1$ actually suggests that the best description
is the M-theory one. In 11 dimensional language, taking into account
the relation $\alpha'=l_p^3/R_{11}$ between the string scale and
the 11 dimensional Planck scale, the masses due to the fluxes are
given by $m \sim l_p^3/R^4$ and are the smallest scale provided that
$R>l_p$, which is the regime where the supergravity approach is
justified.

\section{Discussion}
To sum up our results, we have explored the locus in moduli space around 
$\gs^{-2}g_{ij}=\delta_{ij}$, $A_i=B_{\mu i}=A_{\mu ij}=0$, and probed
how many of the 64 flat directions are lifted by turning on fluxes
on the torus in a supersymmetric way. This requires for instance
that the $H_{ijk}$ and $F_{ijkl}$ fluxes are related by (\ref{sqtorus}).
We have shown that there are ${\cal N}=1$ 
solutions to the equation (\ref{hsl}) that freeze all the moduli that
can possibly be frozen, that is all but a combination of the dilaton
and the volume of the torus. 
The 35 remaining scalars are fixed through a flux-generated potential
that gives them masses when expanding to second order.  
The 28 vectors do not generate scalars
through dualization because the fluxes induce a topological CS mass term
for all of them.
Moreover we have shown that when picking particular configurations
of fluxes, it is possible to produce solutions with all possible
amounts of supersymmetry allowed by the set up, and with different
numbers of moduli. The case where only one massless supermultiplet 
is present does not seem to be generic.

The combination of dilaton and volume which remains massless is such
that large volume implies strong coupling and vice-versa. This makes
it difficult to turn from the purely supergravity point of view that
we consider here to a perturbative string theory approach, since
we would have to deal either with strong coupling or small volumes.
Small volumes would actually imply that the supergravity
approximation breaks down, since non-perturbative objects wrapped
on the torus would be lighter than the lifted moduli.
We are thus driven to consider only the strong coupling, large volume
case.
Hence from the supergravity perspective we could as well consider
that the set up is embedded in 11 dimensions.

We should  also point out the issue of whether any modulus in 3 dimensions
really remains massless after all perturbative corrections are 
taken into account. 
The same issue is relevant for compactifications with
fluxes of M-theory on $Spin(7)$ manifolds.
As discussed in \cite{Acharya} (based on \cite{Shifman}), a massive scalar
in a 3 dimensional ${\cal N}=1$ theory
will receive quantum corrections to its mass at one-loop and 
non-perturbatively. However when the scalar is massless it can be seen
both by standard 3 dimensional arguments (see e.g. \cite{Gates})
and from the 2 dimensional
considerations of \cite{Shifman} that the one-loop correction vanishes.
Thus we are left with the non-perturbative corrections, which typically
will lead to run-away potentials which lift the moduli but bring the
theory to the boundary of the moduli space. The upshot of this is that
it is still interesting, and important, to freeze the moduli at a finite value
by means of a classical potential, generated by the fluxes in the
present case. As for the remaining moduli, the run-away behavior
is potentially disturbing, even if it arises only non-perturbatively.
See however \cite{Louis} for a discussion on how $\alpha'$ corrections
of this type may eventually lead to a stabilization of the remaining
moduli.

Our orientifold example is nothing but a tractable model out of a much
bigger class of generic models, most simply described as M-theory
on $Spin(7)$ manifolds \cite{Gukov,Acharya}. One could in principle apply the
general techniques developed there to our case.
In a similar spirit, it should be possible to make contact between
our configurations with ${\cal N}=2$ supersymmetry and the 
results about compactifications
with fluxes of type II on $G_2$ manifolds \cite{Spence}
or M-theory on Calabi-Yau
fourfolds \cite{GVW,Dasgupta,Haack}. 

\section*{Acknowledgments}
We would like to thank G.~Bonelli, B.~E.~W.~Nilsson and A.~Zaffaroni
for very helpful discussions. R.A.~would like to thank the Physics Department
at Universit\`a di Milano-Bicocca and the Service de Physique
Th\'eorique et Math\'ematique at Universit\'e Libre de Bruxelles
for their hospitality and for providing a very
stimulating environment while this work was completed.
This work is partly supported by EU contract HPRN-CT-2000-00122.

\end{document}